\documentclass{pasa}%
\usepackage{color}

\title[Optimising K dark for KISS]{Optimising the K dark filter for the Kunlun Infrared Sky Survey.}%{Optimising K dark filter for  KISS}
\author[Li et al.]{Yushan Li$^1$, Jessica Zheng$^4$, Peter Tuthill$^1$,  Matthew Freeman$^2$, Michael Ashley$^2$, Michael Burton$^2$, Jon Lawrence$^4$, Jeremy Mould$^3$ \and Lifan Wang$^{5,}$\thanks{Purple Mountain Observatory, Chinese Academy of Sciences, Nanjing, China;
 Chinese Center for Antarctic Astronomy, Nanjing, China}\\
\affil{$^1$University of Sydney}
\affil{$^2$University of New South Wales}
\affil{$^3$Swinburne University}
\affil{$^4$Australian Astronomical Observatory}
\affil{$^5$Texas A\&M University}}

\jid{PASA}
\doi{10.1017/pas.\the\year.\#}
\jyear{\the\year}

\begin{document}

\begin{abstract}
The Kunlun Infrared Sky Survey will be the first comprehensive exploration of the time varying Universe in the infrared.
A key feature in optimizing the scientific yield of this ambitious research program is the choice of the survey passband.
In particular the survey aims to maximally exploit the unique thermal and atmospheric conditions pertaining to the high Antarctic site.
By simulating the expected signal-to-noise for varying filter properties within the so-called ``K$_{DARK}$'' 2.4$\mu$m window, filter performance can be tuned and best-case designs are given covering a range of conditions.  
\end{abstract}

\maketitle

\section{INTRODUCTION }
\label{sec:intro}

\subsection{Scientific goals of KISS}% {\color{red}(Mould and Lifan Wang)} }
The Antarctic plateau is one of  the best sites on Earth for infrared and
submillimeter astronomical observations \cite{lab,yj}. There is an opportunity now to
exploit this scientifically by deploying to Kunlun Station (Dome A) \cite{bcs}
an infrared camera for the AST3-3 wide field telescope.
Location at Kunlun Station offers the supreme advantage of the whole southern sky available for continuous study for the
duration of the Antarctic winter every year. The Antarctic K$_{DARK}$ 2.4$\mu$m passband is a unique 
low background window in the infrared. The Kunlun Infrared Sky Survey (KISS)
 is the first deep and wide K-band survey, and plays a role as a pathfinder for future large antarctic telescopes, such as KDUST \cite{zgb} .

AST3-3 is the third telescope in the AST3 series. Recent results from AST3-1 are reported by \cite{li}.
With its high infrared sensitivity AST3-3 science will emphasize opportunities in the time domain
where longer wavelength measurements are particularly advantageous, such as the following.

\begin{itemize}
\item supernovae and the equation of state
\item reverberation mapping and the physics of AGN
\item Gamma Ray Burster follow up
\item the Cosmic Infrared Background
\item terminal phases of red giants (Miras)
\item dynamics and variability in star formation
\item discovery of exo-planets (esp. brown dwarfs \& hot jupiters)
\end{itemize}

In this paper we consider optimization of the KISS K$_{DARK}$ filter to maximize the signal-to-noise
ratio achievable with this site and instrument. Similar considerations for another exceptional site, 
Mauna Kea Observatory, have been discussed  \cite{st}.
We outline the design of the telescope and camera in section 2 (leaving the details for a further paper),
the Dome A environment in section 3, the optimization algorithm in section 4,
and then draw our conclusions.

\subsection{History of K dark}

The first suggestion that an exceedingly dark window
might exist in the infrared (IR) portion of the spectrum, where the background 
sky brightness could be very much less than at temperate sites, was made by
Harper (1990). This laid the foundations for the development of IR capability at the then fledgling Center for Astrophysical Research in Antarctica (CARA) Dark Sector observatory at the South Pole.

Harper realised that, at the typical --60$^\circ$C winter time temperatures occurring at the South Pole, the thermal background in the K-band (i.e. from 2.0-
2.4$\mu$m) would fall dramatically from the flux at temperate sites, being on the Wien side of the blackbody peak of the sky emission spectrum. Furthermore, between 2.27--2.45$\mu$m there were no known OH airglow lines. Thus, in this narrower band the background flux
could be two orders of magnitude lower than at the best temperate sites, such as Mauna Kea.

From the latter site, observations in K-band are often undertaken in the short portion of the band (i.e. $\lambda~ <$ 2.3$\mu$m), where the emission is dominated by OH
airglow, in order to avoid the steeply rising thermal contribution. %for $\lambda~ <$ 2.3$\mu$m)
In contrast, Harper proposed that from the South Pole sensitivities could be greatly improved by building a telescope that made measurements from 2.27--2.45$\mu$m,
though he did not give this band a name. He even suggested that the observations might be limited by %the flux of 
 the zodiacal light. This is also near its minimum value in this band, between scattered sunlight by
dust in the Ecliptic to shorter wavelengths, and
emission from dust rising to the thermal IR. The gains were quantified in calculations conducted by Lubin (1988) for a
Masters thesis at the University of Chicago; for instance for measurements made by equivalent-sized telescopes
at the South Pole and Mauna Kea, they found that the limiting magnitude may be 2.9 magnitudes fainter at Pole.

The science case for working in this window was then explored (Burton et al. 1993; Burton et al. 1994), as
part of an extensive examination of the potential for
Antarctic observatories across a wide-range of the photon and particle energies. These authors modelled the performance of an Antarctic 2.5m telescope across the
near- and mid-IR bands, as well as the then-planned 8\,m-class telescopes on Mauna Kea (and possible ones in
Australia and Antarctica). These were bench-marked in comparison with the 3.9\,m Anglo Australian Telescope
(AAT). They also took into account the hypothesised ``super-seeing" on the Antarctic plateau (Gillingham
1993). The authors called the 2.27--2.45$\mu$m band K$^+$, and found that an Antarctic 2.5m telescope should be
3.2--3.7~mags more sensitive than the AAT for point
source photometry (the range dependent on the assumptions made), and even 1.2--1.7~mags. more sensitive
than a Mauna Kea 8\,m telescope. However, since for some continuum investigations the precise wavelength
of measurement is not critical, they also compared the results to K$^\prime$ measurements (i.e. 2.0--2.3$\mu$m, without the
thermal longwave end of K-band), finding that there would still be a 0.4~mag. gain over a Mauna Kea 8\,m.

While these predictions were impressive, they lacked site data quantifying the level of the background sky emission
in order to demonstrate their validity, instead relying on
model values. Two experiments were then conducted by CARA at the South Pole to measure these fluxes.
One used the 60 cm SPIREX telescope, equipped with a 1--2.5$\mu$m imaging spectrometer (the GRIM) (Nguyen
et al. 1996). The other used the former 1--5$\mu$m single element photometer/spectrometer from the AAT (the
IRPS) (Ashley et al. 1996), staring at the sky with a tilt mirror through an effective 4$^\circ$ beam. SPIREX
had a specially optimised filter centred at 2.36$\mu$m and running between 2.29 and 2.43$\mu$m, which they named
K$_{DARK}$ -- the first use of this nomenclature. SPIREX found the average value of the zenith sky
flux in K$_{DARK}$ to be 162 $\pm$ 67$\mu$Jy /arcsec$^2$, in contrast to $\sim$ 4000$\mu$Jy /arcsec$^2$ at Mauna Kea. The IRPS
was able to obtain a spectrum of the emission with a resolution of $R = 100$, finding a broad minimum
in sky brightness between 2.30--2.45$\mu$m, and a flux level ranging from 50 to 250$\mu$Jy /arcsec$^2$. Integrated over
the SPIREX K$_{DARK}$ bandpass, the IRPS obtained a mean flux of 180 $\pm$ 60$\mu$Jy /arcsec$^2$, ranging from 60
to 320$\mu$Jy /arcsec$^2$. While exceptionally low, these sky fluxes were still about an order of magnitude higher
than originally predicted by Harper (Harper 1990). Using the measured sky fluxes, the sensitivity gain in the
K$_{DARK}$ band between a 2.5m telescope at the Pole compared to the AAT was then calculated to be a factor of
$\sim$ 50 (Burton 1996).

A more extended analysis of a full winter season of
IRPS data was then undertaken (Phillips et al. 1999).
Between 2.35--2.45$\mu$m the sky brightness  dropped to as low as
50$\mu$Jy /arcsec$^2$, and ranged as high as 200$\mu$Jy arcsec$^2$
in good conditions. A typical value is 100$\mu$Jy /arcsec$^2$ =
17 Vega mags/ arcsec$^2$. No evidence of any auroral contribution to the emission was found. The lowest background
levels were reached once the Sun drops more than 10$^\circ$
below the horizon, when it is no longer able to illuminate the airglow layer (which is at altitudes of 80--100 km). 
However, the lowest flux levels were still about
twice those expected for a 10\% emissive atmosphere at --40$^\circ$C (i.e. as is often found at the top of
the inversion layer). Phillips et al. concluded that the residual emission is probably airglow in origin.

The last IR site testing measurements to have been made in Antarctica come from the NISM instrument,
which was designed as a low-powered version of the IRPS, operating on a power budget of just 10W
(Lawrence et al. 2002). While intended ultimately for deployment to the high plateau, the only measurements
made with the NISM were at South Pole. There was a single filter designed for measuring the K$_{DARK}$ flux
($\lambda_{cent}$ = 2.379$\mu$m $\Delta\lambda$ = 0.226$\mu$m, though with a small
thermal leak which complicated the data analysis). Consistent flux levels were found with the earlier measurements, with a median sky value of $\sim$ 120$\mu$Jy /arcsec$^2$.

The IR fluxes obtained with these three instruments
at South Pole, together with measurements of the superb optical seeing above the surface boundary layer
made at Dome C (Lawrence et al. 2004), formed the basis for the sensitivity calculations made in extensive
science cases (Burton, Storey, \& Ashley 2001; Burton et al. 2005; Lawrence et al. 2009a; Lawrence et al. 2009b;
Lawrence et al. 2009c) for IR telescopes proposed for Dome C. All these science cases suffered, however, from
the lack of any direct measurement of actual sky fluxes from the high plateau.

\subsection{Current NISM experiment}% {\color{red} (Freeman)}}

NISM (Near Infrared Sky Monitor) is the first instrument to measure the infrared sky flux from the high plateau of Antarctica. It was installed at the start of the 2015 season at Ridge A at an elevation of 4040 m. Ridge A is 150 km from Dome A, which is the highest point on the plateau. NISM measures the sky flux using a single pixel InSb diode detector with a cold band-pass filter, which is sensitive to the K$_{DARK}$ window at $2.38 \pm 0.08$ $\mu$m \cite{Bingham2014}.

The NISM detector is pointed at an angled mirror which makes a full 360 degree rotation in elevation once every 10 seconds, sweeping the beam across the sky with a 4 degree field-of-view. In one rotation NISM observes the sky, the ground, and a black body. The black body is a copper cone which is kept at typically 30 degrees above the ambient temperature, and acts as a reference flux to calibrate the instrument. A platinum resistance thermometer measures the black body temperature to an absolute accuracy of better than 0.5$^\circ$C. Further information on NISM can be found in Freeman et al (in preparation).

As these results are not yet available, for the purposes of the present study we will employ the best presently available models and measurements, and allow for several scenarios. Just which of these is most appropriate should be clarified relatively soon with the publication of the NISM results.

\subsection{Motivation: optimising the filter}

To maximise the science yield over the five year KISS survey mission, a key initial design criterion is optimisation of the filter bandpass.
The following sections contain detailed descriptions of the thermal and atmospheric factors relevant to the camera design, our simulation methodology, and finally resultant best-case filters spanning a range of assumptions about the Dome A environment.

\section{Telescope and Camera}

\subsection{Thermal design of the KISS camera}

 For the AST3-3 telescope, Narcissus mirrors \cite{Gillingham} were first proposed to control extraneous thermal backgrounds. As part of the design effort, the thermal emission from telescope and IR camera was investigated more closely, and we examined different thermal control options by either adding a narcissus mirror or a controlling pad behind the telescope secondary mirror. It became clear that a full cold-stop within the camera is a feasible, less risky and economical solution for this kind of survey telescope.

\begin{figure*}
\begin{center}
\includegraphics[width=1.6\columnwidth]{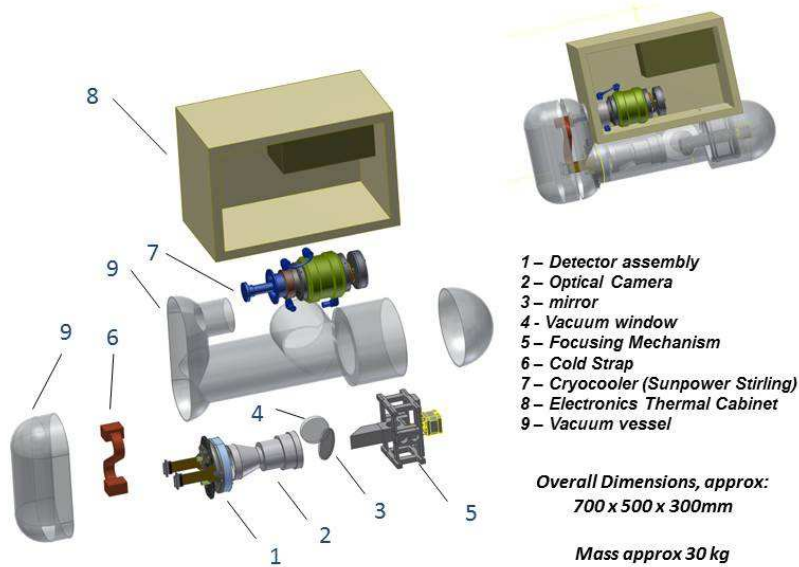}
\caption{KISS camera design.%The detector assembly is coloured  light blue.
%A cold strap is shown in red. The cryocooler and electronics box are the two coloured assemblies towards the top of the figure. 
Light enters the window (4), is reflected from the flat (3) and focusses through the camera (2) on the detector (1).}\label{fig:TelescopeLayout}
\end{center}
\end{figure*}

\subsection{Emissivity of telescope and camera}%

The AST3-3 IR camera is designed for operation at Dome A, Antarctica,  %and works at a wavelength of 2.4$\mu$m 
where the ambient temperature is very low during winter. A major source of background radiation setting the noise floor for observation of faint astronomical sources is black-body thermal emission from the atmosphere and the telescope. These are mostly composed of atmospheric emissivity, emission from telescope mirrors and structures in the beam (such as warm windows), surfaces within the cryostat, as well as thermal emission from within instrument that is scattered onto the detector.  Due to the site's extremely low sky emission (compared to a temperate site), it becomes necessary to perform a rigorous examination of the thermal self emission from the telescope and instrument, as these could play a substantial role in setting the background. 

Figure \ref{fig:TelescopeLayout}  
shows a model of AST3-3 %telescope and 
KISS camera. It% The optical system of the $AST3-3$ 
 includes a transparent entrance window of diameter of 500mm with an Indium Tin Oxide thin film deposited at the front for de-frosting and an aspheric surface at the back for correcting optical aberrations. A primary  and a folding secondary mirror form an f/3.76 intermediate focus before the IR camera re-imaging optics. A field stop is placed here to limit stray light entering the IR camera. The KISS camera includes a flat vacuum window to separate the telescope body from the dewar, and a folding mirror to make the system more compact. The IR re-imaging optics includes four lenses and one filter as well as a cold pupil stop preventing thermal emission from warm camera surfaces arriving on the detector. The final focal ratio of the system is f/5.36, while the plate scale of the IR camera is 1.38$^{\prime\prime}$/pixel (size: 18$\mu$m).  The image quality of the system will be about 1.36$^{\prime\prime}$ providing a match to expectations for episodes of better seeing. %is expected to be of similar order of 04.8$^{\prime\prime}$.

Thermal self-emission (TSE) from structures within telescope and instrument can be calculated from three quantities: (1) the absolute temperature $T$ which determines the spectrum of black body radiation from the Planck function $B_{\lambda}(T)$, (2) the emissivity $\epsilon(\lambda)$ of each component which determines the fraction of black-body radiation's contribution  and (3) the solid angle which is subtended to the detector plane. To estimate the emissivity of optical components, Kirchhoff's law was used. For mirrors, emissivity can be calculated as: $\epsilon  = 1-R$ where $R$ is the measured mirror reflectance. The emissivity of an optical lens can be estimated from its absorption as $\epsilon = absorption $. 

A comprehensive inventory of the properties of opto-mechanical elements making up the telescope and camera was compiled -- key elements such as the tube and baffle of the telescope and KISS camera are painted black with an emissivity of 0.95, while the Indium Tin Oxide coating has emissivity of 0.13. 

Using the FRED\footnote{http://photonengr.com} %\ 
optical engineering simulation package, 
the solid angle from each object was calculated so that the TSE received by detector can be accumulated. 

\begin{figure*}
	\begin{center}
	\includegraphics[width = 1.6\columnwidth]{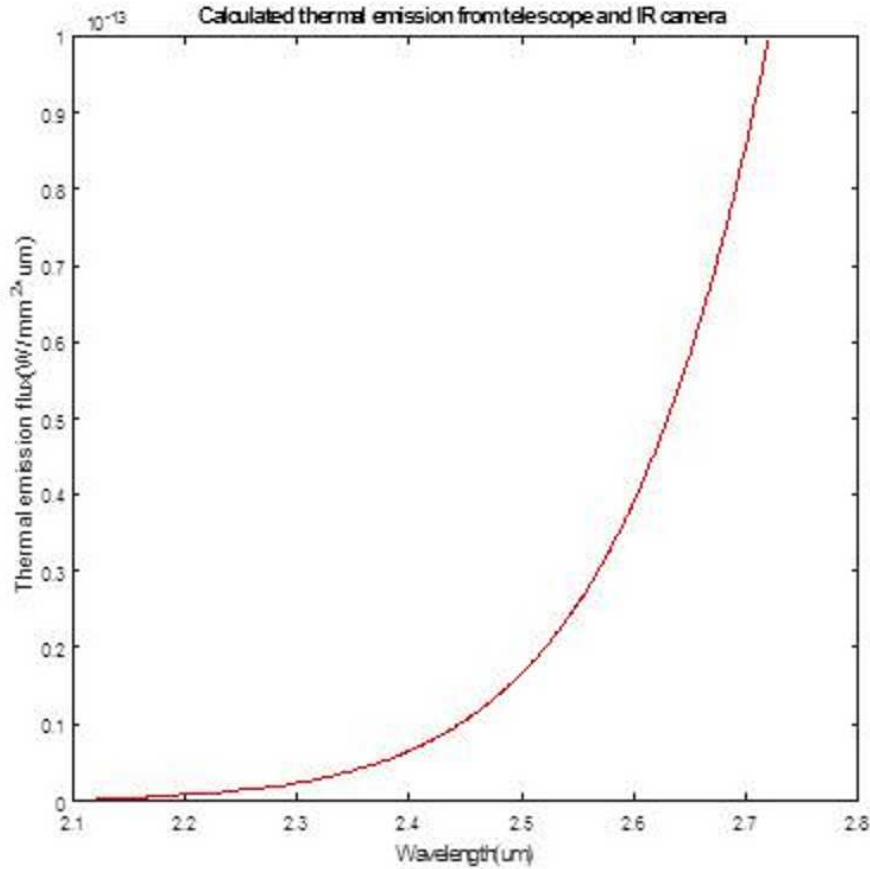}
	\caption{Calculated thermal self-emission on the detector plane }\label{fig:Thermal}
	\end{center}
\end{figure*}

Because the AST-3 telescope was not designed at the outset as an IR optimised instrument, the thermal radiation reaching the detector is actually a factor of a few worse than an idealized design (wherein only unavoidable flux from mirrors and transmissive optics in the optical path contributes).
Mitigating against this, however, is the fact that the telescope enclosure itself is in such a cold environment that the TSE remains lower than the sky contribution. 
The solid angle for each object is then examined using a Matlab program and by varying the temperature of each object according to its position within the telescope, the total TSE can be obtained. Figure \ref{fig:Thermal} shows the calculated thermal emission radiance from the system, when the telescope ambient temperature is 210K, and the IR camera optics chamber is 150K and the detector is 77K.

\section{The Atmosphere and Environment}

\subsection{Temperature range of Dome A and ground level seeing}

There are, as yet, no direct measurements of the night time seeing at Dome A.  However, and as discussed in Burton et al.\ (2010), we may anticipate the value for the seeing at Dome A based on measurements of meteorological parameters, and by comparison with the levels determined at other high plateau sites, in particular the South Pole and Dome A\@.

There is a narrow and intense boundary layer above Dome A\@.  Measurements of its height have been made using a sonic radar (a SNODAR) \cite{2010PASP..122.1122B}, yielding a median value of 13.9\,m, and a 25\% quartile of just 9.7\,m.  For a telescope placed above this boundary layer it is anticipated that it will experience only the free atmosphere seeing.  The value should be at least as good as that measured from Dome C (which is 900\,m lower in elevation); here the median value was found to be $0.27''$ in V--band \cite{2004Natur.431..278L}.  The boundary layer at Dome A is also much thinner than at the South Pole, where it is $\sim 200$\,m \cite{2002A&A...385..328M}, and about half that at Dome C (23-27\,m) \cite{2009A&A...499..955A}, so facilitating the ease of raising a telescope above it.

Within the boundary layer there is a strong temperature gradient.  Analysis of temperature measurements made from a weather tower at Dome A \cite{Hu} show that typically this is $15^\circ$C between the ice surface of the top of the boundary layer; i.e.\ a temperature gradient of $\sim \rm 1^{\circ}C/m$.  The gradient is particularly intense close to the surface; for instance the mean temperatures measured by Hu et al.\ at 0, 2 \& 4\,m above the ice are $\rm -60^{\circ}C, -54^{\circ}C \, \& -52^{\circ}$C, respectively. 

There is also a considerable range in the temperature at Dome A.  While hourly variations are generally less than $1^{\circ}$C (and there are no diurnal variations in winter), over week long periods the temperature can vary by up to $20^{\circ}$C\@.  For instance, over the winter period, and at a height 4\,m above the ice, the temperature ranges from $-70^{\circ}$ to $-45^{\circ}$C. 
Within the boundary layer the contribution to the seeing is considerable.  At any given height close to the ice the seeing is also variable, depending particularly on whether it happens to be above the boundary layer at the time.  We estimate the boundary layer seeing at Dome A by comparison with measurements made at other high plateau sites.  

At the South Pole this layer contributes typically $1.5''$ at Pole in V--band \cite{1996A&AS..118..385M, 1999A&AS..134..161M} (and $1.1''$ at 2.4$\mu$m) \cite{2002A&A...385..328M}.  
The distribution of seeing measurements close to the ice surface at Dome C has a bi-modal distribution \cite{2009A&A...499..955A}; at 8\,m elevation the two peaks are at $0.4''$ and $1.6''$.  The former corresponds to times when the top of boundary layer drops below 8\,m and occurs about 20\% of the time.  The latter is the more normal condition, with the mean boundary layer at Dome C being determined by Aristidi et al.\ to be between 23--27\,m.  
At Dome F, daytime measurements made from an instrument placed 11\,m above the ice \cite{2013A&A...554L...5O} found the seeing to sometimes drop below $0.2''$ at V--band.  This would then correspond to the free-atmosphere value, with the monitor then above the boundary layer. A median value was found to be $0.54''$, with a tail extending up to $3''$.  These include the contribution from the free atmosphere as well as the boundary layer, whose median height is estimated to be 18\,m.

We anticipate a similar situation to also occur at Dome A, compressed into the $\sim 15$\,m depth of the boundary layer.  The boundary layer seeing would then be $\sim 1''$ at $\rm K_{dark}$, and when the top boundary layer drops below the height of a telescope, the free atmosphere seeing of better than $0.3''$ would be obtained.  Direct measurements of the seeing at Dome A are desirable.

\subsection{Atmospheric Transmission}

Unfortunately, dedicated measurements of the atmospheric transmission spanning the near-infrared 
are not available for the Dome~A site. 
Summaries of the status of Antarctic observational campaigns have been given  \cite{1998ASPC..141....3B}.

However a sophisticated treatment employing the atmospheric modelling program MODTRAN to model the
observed sky spectrum at the South Pole from the near-IR to the sub-mm has been used \cite{hd}.
The model incorporates more than a dozen atomic and molecular species, as well as the 
effects of aerosols (but does not incorporate a contribution from airglow). 
Extracted below in Figure~\ref{fig:transmission} are models of the polar atmospheric transmission
over the two micron region. 
These span a range of values for the total precipitable water content with the ``mid'' model corresponding
to 164\,$\mu$m, while ``wet'' and ``dry'' are 324\,$\mu$m and 82\,$\mu$m respectively.

\begin{figure*}
	\begin{center}
	\includegraphics[width = 1.9\columnwidth]{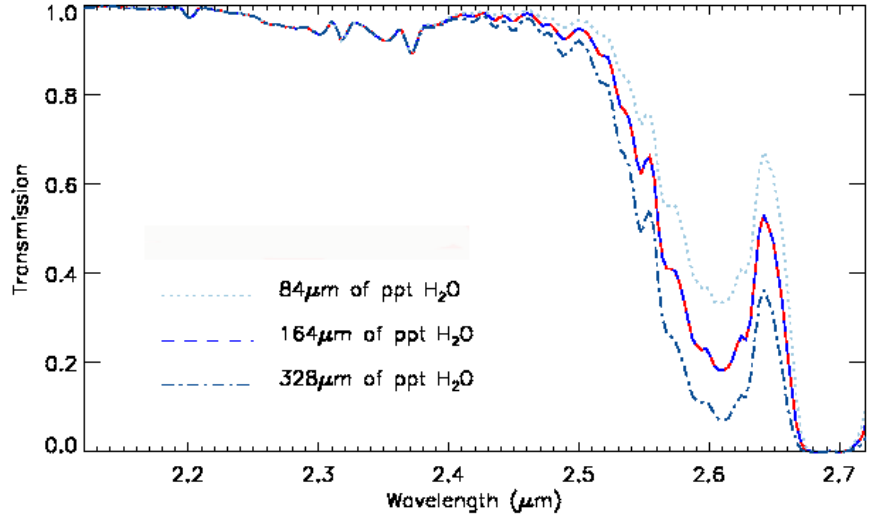}
        \vspace*{-3 truein}
	\caption{Curves of South Pole atmospheric transmission taken from Hidas et al. 2000, 
                with varying assumptions about the atmospheric water content.% as described in the text.
The three lines are for precipitable water columns  of 84$\mu$m, 164$\mu$m and 328$\mu$m.
%                For convenience also overplotted here is the expected loss term due to the declining 
%                Quantum Efficiency of the detector into the long wavelength infrared.
}
        \label{fig:transmission}
	\end{center}
\end{figure*}

\subsection{Atmospheric OH}

\begin{figure*}
\begin{center}
\includegraphics[width=\columnwidth,angle=-90]{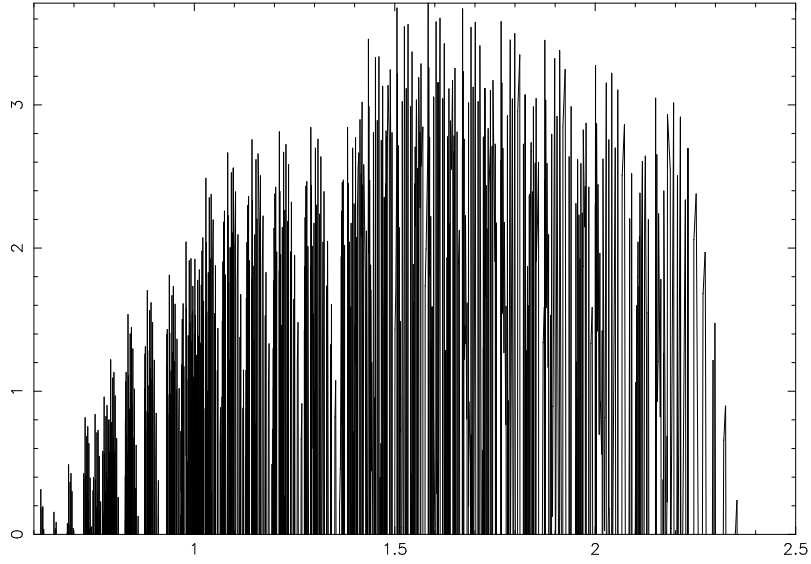}
\caption{Intensity of OH lines in the catalog of Rousselot $et~al.$ (log scale).}\label{fig:Fig1a}
\end{center}
\end{figure*}

\begin{figure*}
\begin{center}
\includegraphics[width=\columnwidth,angle=-90]{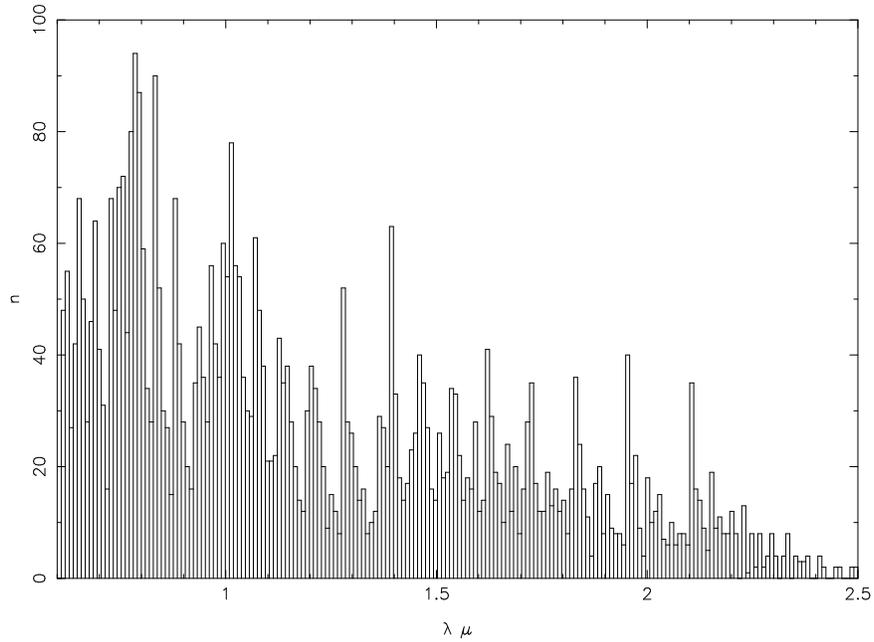}
\caption{Number of lines in the catalog. The horizontal axis is wavelength in this and the previous figure. }\label{fig:Fig1b}
\end{center}
\end{figure*}

One inescapable source of infrared background for observations from ground based observatories
is the vibration-rotation spectrum of the OH molecule. The first overtone band
is centred at 1.6$\mu$m, but the higher vibrational states emit in lines
that extend into the K band window. Highly useful data are available \cite{rou} , shown here as Figure  \ref{fig:Fig1a}
\& Figure \ref{fig:Fig1b}. Provided the short wavelength cutoff $\lambda_1~>$ 2.3$\mu$m, the background variability due to airglow will be minimized.

\subsection{Thermal background emissivity and the atmosphere}

\begin{figure*}
	\begin{center}
		\includegraphics[width=1.9\columnwidth]{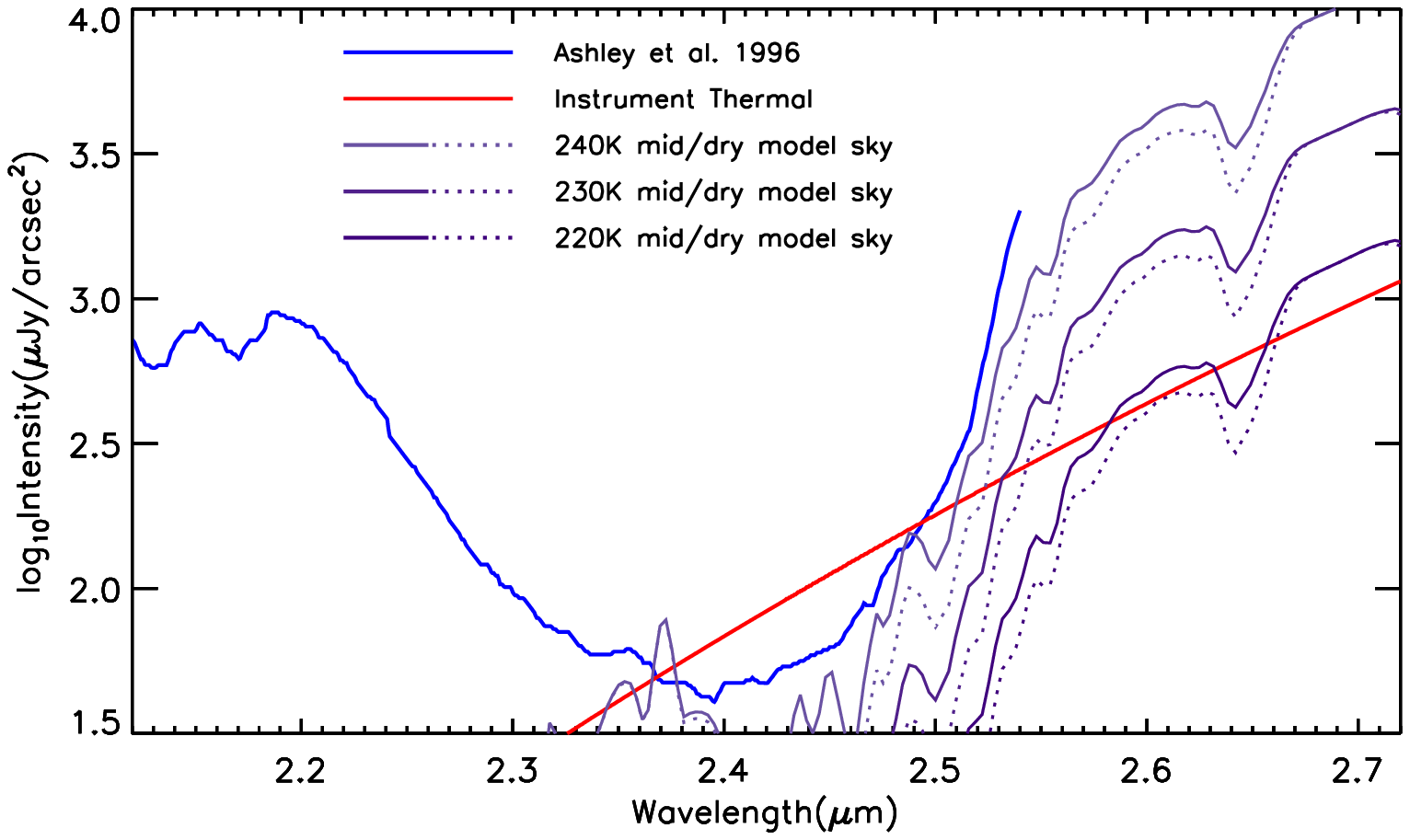}
                \vspace*{-3 truein}
		\caption{Measured infrared sky brightness from the South Pole taken from Ashley et al. (1996).% and Phillips et al. 1999.
                Thermal self-emission from the instrumentation, as well as simple emissivity models assuming a cold, dry 
                dome-A atmosphere are also presented (further details in the text). Note that Figure 1 of Phillips et al. (1999) provides additional observations of the sky from the South Pole, however these data appear to have an error in their wavelength calibration, and so are not used here.} 
        \label{fig:skybright}
	\end{center}
\end{figure*}

Measurements of the thermal sky background are available \cite{a96} and \cite{1999ApJ...527.1009P}.
These pertain to the South Pole, and conditions on the plateau are expected
to be significantly darker. The measured sky brightness from these two sources has been plotted in Figure~\ref{fig:skybright}.
Note that these two sources yield somewhat discrepant data, and in particular both the position and the shape of the K$_{DARK}$ region
around 2.4\,$\mu$m differing. For the purposes of optimisation of the KISS survey bandpass in later 
sections, we have adopted the %\cite{a96} %
Ashley et al. (1996) 
curves (in which the 2.4\,$\mu$m dark region plays a more accentuated role).
For the purposes of optimising the shape of the $\rm K_{DARK}$ bandpass we make use of the sky spectrum presented in %\cite{a96}%
Ashley et al. (1996) 
as opposed to that in Phillips et al. (1999).  The former represents the best conditions at South Pole, whereas the latter the average conditions.  We expect it to provide a better model for the colder and clearer conditions of Dome A compared to South Pole.

Also overplotted in Figure~\ref{fig:skybright} is the expected thermal background arising from the camera and telescope structure itself. 
It is worth noting that the instrumental self-radiation exceeds the
sky background thermal noise floor (assuming  Ashley et al. 1996) particularly
towards the red end of the spectral range, and therefore future instruments with more dedicated and optimised infrared design stand to make further
sensitivity gains. 

Because all observational data available pertain to the South Pole while the high plateau is expected to be significantly colder and drier, we also 
overplot in Figure~\ref{fig:skybright} some very simple thermal atmosphere  model predictions.
These were obtained by assuming that the atmospheric emissivity is the
converse of the transmission as given by the MODTRAN models discussed above
and presented in Figure~\ref{fig:transmission}. Predictions for the ``red edge" of the K$_{DARK}$ window are seen to shift to 
longer wavelength as the assumed effective temperature of the atmosphere
becomes colder. Note that this simple model does not include any airglow component, and so
loses predictive power at the blue side of the band. 

\section{Algorithm}

\subsection{Analytical Model}

For the purposes of the optimisation presented here, a simplified model for the background noise processes on the detector has been adopted.
%In particular, readout noise and other electronic or detector noise processes have been neglected, 
The following noise sources (all assumed Poisson distributed) are considered to contribute to the signal-to-noise calculation:

\begin{itemize}
	\item photon noise on the signal($S$) from the object
	\item photon noise on the background flux ($B$) from the sky
	\item photon noise on the thermal self emission (TSE) from telescope and IR camera
	\item read out noise, $r$ electrons
	\item shot noise on the dark current ($I_{D}$) 
\end{itemize}

The signal to noise ratio can found from the equation: 
\begin{equation} \label{SN}
\frac{S}{N} = \frac{S \sqrt{t}} {\sqrt{S + n_{p}(B+TSE+I_{D}+\frac{r^{2}}{t})}}
%\frac{S}{N} = \frac{S \sqrt{t}}{\sqrt{S + n_{p}(B+TSE+I_{D}+r/t)}}
\end{equation}

 where $t$ is the observing integration time, %and $n_{0}$ can be replaced with 1 when only one exposure is used. 
$n_{p}$ is the number of pixels sampling the stellar image and can be estimated by:

\begin{align}
 n_{p} = \frac{\pi}{4}(\frac{\theta_{FWHM}} {\theta_{p}})^2 \\
  \theta_{FWHM} = 2.35\sigma \\
  \sigma  = \sqrt{({\rm seeing}^2 + {\rm Image~quality}^2 )}
\end{align}
where $\sigma$ is the image quality dictated by the seeing and the point spread function(PSF) quantified as the 1$\sigma$ width of a Gaussian beam. %(1.38'' for AST3-3 IR camera). 
$\theta_{FWHM}$ is the Full Width of Half Maximum intensity of the PSF, and  $\theta_{p}$ is the pixel scale of the system.
The image quality of 1.36$''$ is from the diffracted image size in combination of seeing 0.6$''$ and a system tolerance of 0.62$''$.
The photons collected by a telescope of area $A$    %_{tel}$ 
in a wavelength range of $\lambda$ from an object of apparent magnitude of $m$, transmitted through the atmosphere with transmission of $T_{Sky}(\lambda)$, then pass through an optical system of efficiency $\tau(\lambda)$ on to a CCD detector with quantum efficiency of $\eta(\lambda)$ and with a filter response of $T_{f}(\lambda) $, can be calculated as:

\begin{align}
 S = \int _{\lambda1} ^{\lambda2} F_{\lambda}(0)10^{-0.4 m} A T_{Sky}(\lambda) \tau(\lambda) \eta(\lambda)T_{f}(\lambda) \lambda (hc)^{-1}d\lambda
 %S = \int _{\lambda1} ^{\lambda2} F_{\lambda}(0) 10^{-0.4 m} A_{tel} T_{Sky}(\lambda) \tau(\lambda) \eta(\lambda) \\*T_{filter}(\lambda) \lambda (hc)^{-1}d\lambda
\end{align}
where $F_{\lambda}(0)$ is the flux in $ \rm{W ~cm^{-2} \mu m^{-1}}$ from a zeroth magnitude standard star above the atmosphere.

Similar methods can be used to estimate the sky background noise $B$   and thermal self-emission from telescope TSE received by one pixel as described in \cite{McLean} and they are represented  in equation~\ref{Background}. %and \ref{Thermal}.

\begin{align} \label{Background}
B = \int_{\lambda1} ^{\lambda2}F_{\lambda}(0) 10^{-0.4m_{sky}(\lambda)}A\tau(\lambda)\eta(\lambda)T_{f}(\lambda)\theta_{p}^{2}\lambda(hc)^{-1}d\lambda 
%   B = \int _{\lambda1} ^{\lambda2}  F_{\lambda}(0) 10^{-0.4m_{sky}(\lambda)} A_{tel}  \tau(\lambda) \eta(\lambda)\\T_{filter}(\lambda) \theta_{pixel} ^{2} \lambda (hc)^{-1}d\lambda 
\end{align}

 It is obvious that the signal to noise ratio (SNR) is related to the spectrum properties of all parts in the optical path. When a system is designed,  normally the spectrum of the sky transmission, sky emission, the quantum efficiency of the CCD detector is defined. The thermal self emission from the telescope could be varied with the ambient temperature. In order to obtain the best signal to noise ratio for an observed object, the  filter spectral response needs to be  carefully designed. 

\section{Results}

\subsection{The optimal KISS observing bandpass}

%Having defined the signal-to-noise ratio (SNR) in the previous section, 
It is now  relatively straightforward to compute the SNR while varying a wide range of environmental and observing conditions. 
Large SNR grid calculations optimised the two key filter parameters: the bandwidth and center wavelength of the filter; while also spanning ranges of parameters such as the thermal environment, seeing, brightness of the target star.
For the purposes of the calculations, a square response filter with some rounding at the edges as is typical for multi-layered interference filters was assumed.

Assuming the atmospheric thermal spectrum at the site to follow that published in Ashley et al. 1996 (for the Pole), then an optimal KISS survey filter was found to have a centre frequency and bandpass of  2.36\,$\mu$m  and 0.235\,$\mu$m respectively.
The gridded SNR data illustrating this optimisation run is given in Figure~\ref{fig:bestfilter}.

\begin{figure*}
	\begin{center}  
        \label{fig:bestfilter}
	\includegraphics[angle=90, width = 1.9\columnwidth]{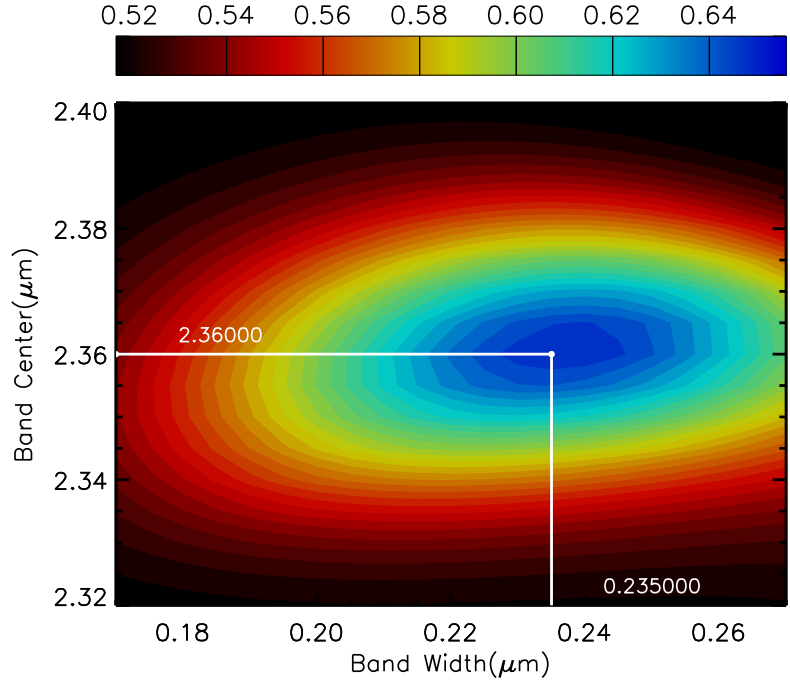}
        \vspace*{-3 truein}          
	\caption{Contours produced by the optimisation algorithm exploring KISS survey SNR as a function of the bandwidth and centre wavelength of the chosen filter. 
                The peak denoting an optimum filter for this case is marked. }  \label{fig:bestfilter}
	\end{center}
\end{figure*}

The calculations optimising the SNR were found to yield a fairly robust outcome even under changing conditions. 
Increasing the thermal ambient environment acted, as would be expected, to shorten the red edge of the filter bandpass. 
The effect of varying target brightness was also explored, and found to have little effect on the optimisation except for cases of good ($\sim$~1$''$) seeing.
Then only the filter bandwidth (not centre wavelength) was seen to increase by a modest $\sim$10\% as the flux of the target star was increased from 16th mag to 14th mag. Further simulations with the input blackbody varying from 2000K to 10000K show that the star's spectrum only slightly modifies the optimized filter curve. The optimized center wavelength of the filter remains within 0.01$\mu$m of nominal value of 2.375$\mu$m and the bandwidth within 0.02$\mu$m of nominal.

\subsection{Influence of the sky background on the optimal bandpass}

As illustrated in Figure~\ref{fig:skybright}, expectations for the red edge of the sky thermal background may be significantly pessimistic when applying data taken at the South Pole to expectations for the high plateau.
As described above, we have formulated a simple model that might predict the shift of this emission edge with the expected colder and drier environoment at Dome~A. 
Over the assumed range of atmospheric conditions explored earlier, Table~\ref{tab:colddry_atm} gives the  optimum filter properties for each model atmosphere.
Moderate shifts at the 10\% level towards the red are witnessed as expectations for the atmospheric background become colder and drier.

\begin{table}
\caption{Optimum filter properties (centre and bandpass in microns) as a function of atmospheric properties.}
\begin{center}
\begin{tabular}{ccc}
\hline\hline
Atm Temp & ``mid" 164\,$\mu$m PWV & ``dry" 82\,$\mu$m PWV \\
\hline%
240K  &      2.365, 0.240   &     2.370, 0.245 \\
230K  &      2.375, 0.250   &     2.375, 0.255 \\
220K  &      2.380, 0.255   &     2.380, 0.260 \\
\hline
\end{tabular}
\end{center}
\label{tab:colddry_atm}
\end{table}

 \begin{figure*}
	 \begin{center}
		 \includegraphics[width = 1.9\columnwidth]{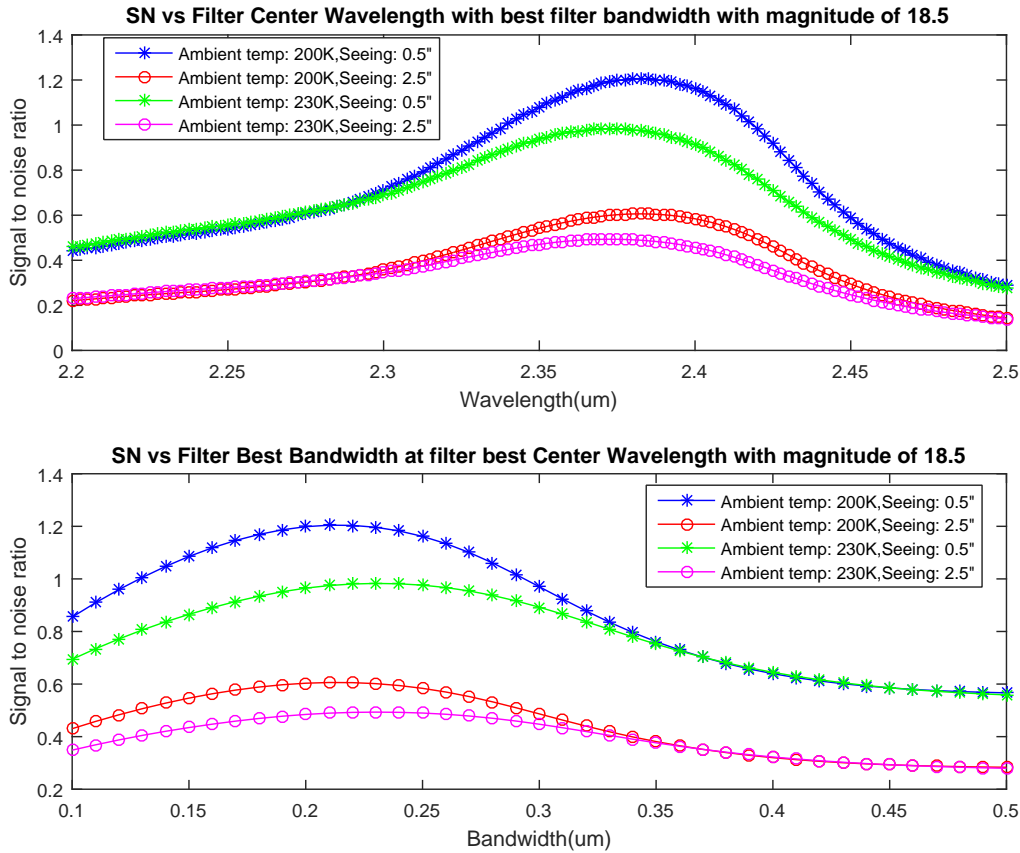}
		 \caption{Filter SNR vs bandwidth and centre wavelength.}% (a) Magnitude of observed object:18.5. (b)Magnitude of observed object:14 }
\label{fig:FilterContour}
	 \end{center}
 \end{figure*}

 \subsection{Effect of filter wings} %using NISM filter wings as typical

The exponential rise in background in Figure 2 implies that we need to fully suppress a red leak from the K$_{DARK}$ %the long wavelength 
 filter wings. \cite{st} note that
 the manufacturing specifications of their filters were designed so that out of band transmission $<$0.0001 out to 5.6 $\mu$m
%    All parameters for 65 K; cold filter scans of witness samples to be provided with prediction of wavelength shift with temperature.(*)
%    >80% average transmission (goal >90%).
%    Peak transmission level of broad band filters to have a ripple of less than ±5% of average transmission between 80% points.
 and   cut-on and cut-off wavelengths are to be attained within $\pm$0.5\%. The KISS detector QE is specified to exceed 50\% at 2.5$\mu$m but falls to zero at
approximately 3$\mu$m.
The filters of \cite{st} have a
roll-off spec: slope less than or equal to 2.5\% where slope is defined to be: [ $\lambda$(80\%)--$\lambda$(5\%) ] / $\lambda$(5\%). 
This is a steeper slope beyond 2.5$\mu$m than the rise in Figure 2.

\section{Conclusions}

A realistic simulation of the performance of the instrumentation responsible for performing the KISS survey has been undertaken, exploring a range of possible environmental conditions and using the best currently available Antarctic site-testing data.
The parameters of the bandpass filter to best exploit the previously reported K$_{DARK}$ window have been examined. 
These were found to vary only modestly over the range of likely assumptions for the environment and observing conditions, enabling a reasonably firm specification ($\lambda_0$ 2.375, $\Delta\lambda$ 0.25$\mu$m) to be made in furnishing the survey optical hardware.

\begin{acknowledgements}
We would like to thank all our colleagues on the KISS team for helpful discussions.
The KISS camera is acquired through Australian Research Council LIEF grant LE15100024.

\end{acknowledgements}

% UNCOMMENT THE LINES BELOW IF YOU WISH TO USE BIBTEX
%\bibliographystyle{apj}
%\bibliography{yourbibfile}

\begin{thebibliography}{}

\bibitem[Aristidi et al 2009]{2009A&A...499..955A}	      
	Aristidi, E.; Fossat, E.; Agabi, A.; Mekarnia, D.; Jeanneaux, F.; Bondoux, E.; Challita, Z.; Ziad, A.; Vernin, J.; Trinquet, H.
2009, A\&A, 499, 955

\bibitem[Ashley et al. 1996]{a96} Ashley, M.~C.~B., Burton, M.~G., Storey, J.~W.~V., et al.\ 1996, PASP, 108, 721 

\bibitem[\protect\citename{Bingham and Ashley~}2014]{Bingham2014}
Bingham, N. R., Ashley, M. C. B., 2014, 
Society of Photo-Optical Instrumentation Engineers (SPIE) Conference Series, 9154, 91541V

\bibitem[Bonner et al. 2010]{2010PASP..122.1122B}	      
	Bonner, C. S.; Ashley, M. C. B.; Cui, X.; Feng, L.; Gong, X.; Lawrence, J. S.; Luong-van, D. M.; Shang, Z.; Storey, J. W. V.; Wang, L. et al 2010,
PASP, 122, 1122

\bibitem[Burton 1996]{two} Burton, M.~G.\ 1996, PASA, 13, 2 
\bibitem[Burton 1998]{1998ASPC..141....3B}      
	Burton, M.  1998, ASP Conf Ser 141, 3
\bibitem[Burton, Storey, \& Ashley 2001]{four} Burton M.~G., Storey J.~W.~V., Ashley M.~C.~B., 2001, PASA, 18, 158 
\bibitem[Burton et al. 2005]{five} Burton M.~G., et al., 2005, PASA, 22, 199 
\bibitem[Burton et al 2010]{six}                          
	Burton, M. G.; Burgarella, D.; Andersen, M.; Busso, M.; Eiroa, C.; Epchtein, N.; Maillard, J.-P.; Persi, P. 2010 EAS, 40, 125
\bibitem[Gillingham 1993]{prg} Gillingham, P.R. 1993,  ANARE Research Notes 88, 290. Burns G., Duldig M. (eds) 

\bibitem[\protect\citename{Gillingham}~2002]{Gillingham} Gillingham, P.R. 2002, PASA, 19, 301

\bibitem[Harper 1990]{1990AIPC..198..123H} Harper, D.~A.\ 1990, American Institute of Physics Conference Series, 198, 123 

\bibitem[Hidas et al. 2000]{hd}
Hidas, M. G.; Burton, M. G.; Chamberlain, M. A.; Storey, J. W. V. 2000, PASA, 17, 260	

\bibitem[Hu et al 2014]{Hu}	
	Hu, Yi; Shang, Zhaohui; Ashley, Michael C. B.; Bonner, Colin S.; Hu, Keliang; Liu, Qiang; Li, Yuansheng; Ma, Bin; Wang, Lifan; Wen, Haikun 2014,
	PASP, 126, 868-881 

\bibitem[Lawrence et al. 2002]{2002PASA...19..328L} Lawrence, J.~S., Ashley, M.~C.~B., Burton, M.~G., et al.\ 2002, PASA, 19, 328 

\bibitem[Lawrence et al. 2004]{2004Natur.431..278L} Lawrence J.~S., Ashley M.~C.~B., Tokovinin A., Travouillon T., 2004, Nature, 431, 278 

\bibitem[\protect\citename{Lawrence et al.~}2006]{lab}Lawrence, J. S. et al. 2006, SPIE 6267, 1

\bibitem[Lawrence et al. 2009a]{2009PASA...26..379L} Lawrence J.~S., et al., 2009, PASA, 26, 379 

\bibitem[Lawrence et al. 2009b]{2009PASA...26..397L} Lawrence, J.~S., Ashley, M.~C.~B., Bunker, A., et al.\ 2009, PASA, 26, 397 

\bibitem[Lawrence et al. 2009c]{2009PASA...26..415L} Lawrence, J.~S., Ashley, M.~C.~B., Bailey, J., et al.\ 2009, PASA, 26, 415 

%Lawrence, J. S.; Ashley, M. C. B.; Burton, M. G.; Cui, X.; Everett, J. R.; Indermuehle, B. T.; Kenyon, S. L.; Luong-Van, D.; Moore, A. M.; Storey, J. W. V.; and 5 coauthors 2006 SPIE 6267, 1

\bibitem[Li et al. 2015] {li}
Li, G., Fu, J.N. \& Liu, X.M. 2015, arxiv 1510.06134

\bibitem[Lubin 1988]{lubin} Lubin, D.\ 1988, Masters Thesis, Univ. Chicago

\bibitem[Marks 2002]{2002A&A...385..328M}	      
	Marks, R. D. 2002, A\&A, 385, 328

\bibitem[Marks et al 1996]{1996A&AS..118..385M}      
	Marks, R. D.; Vernin, J.; Azouit, M.; Briggs, J. W.; Burton, M. G.; Ashley, M. C. B.; Manigault, J. F. 1996, A\&AS 118, 385

\bibitem[Marks et al 1999]{1999A&AS..134..161M}	     
	Marks, R. D.; Vernin, J.; Azouit, M.; Manigault, J. F.; Clevelin, C. 1999, A\&AS, 134, 161

\bibitem[McLean 2008]{McLean}
 McLean: Ian S. 2008;  Electronic Imaging in Astronomy Detectors and Instrumentation(2nd edn; Chichester,UK Springer)
%Meteorological Data for the Astronomical Site at Dome A, Antarctica

\bibitem[Nguyen et al. 1996]{1996PASP..108..718N} Nguyen, H.~T., Rauscher, B.~J., Severson, S.~A., et al.\ 1996, PASP, 108, 718 

\bibitem[Okita et al 2013]{2013A&A...554L...5O}
	Okita, H.; Ichikawa, T.; Ashley, M. C. B.; Takato, N.; Motoyama, H. 2013, A\&A, 554, 50

\bibitem[Phillips et al. 1999]{1999ApJ...527.1009P} Phillips, A., Burton, M.~G., Ashley, M.~C.~B., et al.\ 1999, ApJ, 527, 1009 

\bibitem[\protect\citename{Rousselot et al. ~}2000]{rou}
Rousselot, P., Lidman, C., Cuby, J.-G., Moreels, G., Monnet, G. 2000, A\&A, 354, 1134
                                                                       
\bibitem[Simons \& Tokunaga 2002]{st}  
Simons, D. A. \& Tokunaga, A. 2002, PASP, 114, 169	

\bibitem[\protect\citename{Ji Yang et al.~ }2010]{yj}                          
%Yang, Ji; Zuo, Ying-Xi; Lou, Zheng; Cheng, Jing-Quan; Zhang, Qi-Zhou; Shi, Sheng-Cai; Huang, Jia-Sheng; Yao, Qi-Jun; Wang, Zhong 2013, RAA, 13, 1493
Yang, Ji et al. 2013, RAA, 13, 1493

\bibitem[\protect\citename{Yuan et al.~}2015]{bcs}
%Yuan, Xiangyan; Cui, Xiangqun; Wang, Lifan; Gu, Bozhong; Du, Fujia; Li, Zhengyang; Yang, Shihai; Li, Xiaoyan; Lu, Haiping; Wen, Haikun; and 3 coauthors 2015 IAUGA, 225, 6923
Yuan, Xiangyan et al 2015, IAUGA, 225, 6923

\bibitem[\protect\citename{Zhao et al. }2011]{zgb}
%Zhao, Gong-Bo; Zhan, Hu; Wang, Lifan; Fan, Zuhui; Zhang, Xinmin 2011, PASP, 123, 725
Zhao, Gong-Bo et al. 2011, PASP, 123, 725

%\bibitem[Burton et al. 1994]{three} Burton, M., Aitken, D.~K., Allen, D.~A., et al.\ 1994, PASA, 11, 127 
%\bibitem[Burton et al. 1993]{bam} Burton M.G., Allen D.A., McGregor P.J. 1993,  ANARE Research Notes 88, 293. Burns G., Duldig M. (eds) 
%\bibitem[Ashley et al.(1995)]{1995SPIE.2552...33A} Ashley, M.~C., Burton, M.~G., Lloyd, J.~P., \& Storey, J.~W.\ 1995, Proc. SPIE, 2552, 33 
%\bibitem[Burton et al.(1999)]{1999ldss.work..201B} Burton, M.~G., Storey, J.~W.~V., \& Ashley, M.~C.~B.\ 1999, Looking Deep in the Southern Sky, 201 
%\bibitem[Harper(1994)]{1994ASSL..190..247H} Harper, D.~A.\ 1994, Astronomy with Arrays, The Next Generation, 190, 247 
%\bibitem[Rauscher et al. 1995]{1995ASPC...79..195R} Rauscher, B.~J., Hereld, M., Nguyen, H., \& Severson, S.\ 1995, Robotic Telescopes.~ Current Capabilities, Present Developments, and Future Prospects for Automated Astronomy, 79, 195 
%\bibitem[Storey et al. 1996]{1996PASA...13...35S} Storey, J.~W.~V., Ashley, M.~C.~B., \& Burton, M.~G.\ 1996, PASA, 13, 35 

\end{thebibliography}

%\bibitem[\protect\citename{Ashley et al.~ }1996]{a96}     
%	Ashley, Michael C. B.; Burton, Michael G.; Storey, John W. V.; Lloyd, James P.; Bally, John; Briggs, John; Harper, Doyal A. 1996, PASP, 108, 721
%	Ashley, Michael C. B. et al.~ 1996, PASP, 108, 721
%Burton M.G., Allen D.A., McGregor P.J. 1993, ANARE Research Notes 88, 293. Burns G., Duldig M. (eds)\\
%Hereld, M. 1994, Experimental Astronomy, 3, 87\\

\end{document}